\documentclass[useAMS,usenatbib]{mn2e}
\usepackage[hidelinks]{hyperref}
\usepackage{graphicx}
\usepackage{float}
\usepackage{amsmath}
\usepackage{mathtools}
\usepackage{amssymb}
\usepackage{cite}
\usepackage{booktabs} 
\usepackage{multirow} 
\usepackage{url} 

\title[The SPINN classifier]{SPINN: a straightforward machine learning solution to the pulsar candidate selection problem}
\date{June 13, 2014}
\author[V.~Morello et al.]
	{V.~Morello$^{1,2}$, E.D.~Barr$^{1,2}$, M.~Bailes$^{1,2}$, C.M.~Flynn$^{1}$, E.F.~Keane$^{1,2}$
\newauthor and W.~van Straten$^{1,2}$\\
	$^1$Centre for Astrophysics and Supercomputing, Swinburne University of Technology, PO Box 218, VIC 3122, Australia \\
	$^2$The ARC Centre of Excellence for All-Sky Astrophysics (CAASTRO)
	}

\begin{document}

\maketitle

\begin{abstract}
	We describe SPINN (Straightforward Pulsar Identification using Neural Networks), a high-performance machine learning solution developed to process increasingly large data outputs from pulsar surveys. SPINN has been cross-validated on candidates from the southern High Time Resolution Universe (HTRU) survey and shown to identify every known pulsar found in the survey data while maintaining a false positive rate of 0.64\%. Furthermore, it ranks 99\% of pulsars among the top 0.11\% of candidates, and 95\% among the top 0.01\%. In conjunction with the \textsc{peasoup} pipeline \citep{Barr2014}, it has already discovered four new pulsars in a re-processing of the intermediate Galactic latitude area of HTRU, three of which have spin periods shorter than 5 milliseconds. SPINN's ability to reduce the amount of candidates to visually inspect by up to four orders of magnitude makes it a very promising tool for future large-scale pulsar surveys. In an effort to provide a common testing ground for pulsar candidate selection tools and stimulate interest in their development, we also make publicly available the set of candidates on which SPINN was cross-validated.
\end{abstract}

\section{Introduction}
	Discovering pulsars typically involves identifying periodic signals in observational data, then reducing each of them into a set of diagnostic values and graphical representations referred to as a \textit{candidate}. A modern all-sky pulsar survey such as the High Time Resolution Universe \citep[HTRU;][]{Keith2010} produces several million such candidates, the overwhelming majority of which are either the result of human-made radio-frequency interference (RFI), or due to various forms of noise. The selection of promising candidates to be observed again for confirmation remains up to this day heavily dependent on human inspection, a very time-consuming process becoming increasingly unmanageable as surveys continue to evolve into ever larger scale operations over time.
	Next generation instruments such as the Square Kilometre Array (SKA), can be expected to find 20,000 pulsars \citep{Smits2009}, but not before an estimated 200 million candidates are properly classified, if we are to conservatively assume that the fraction of pulsars to be found among them (one in ten thousand at most) is comparable to current surveys \citep{Lyon2013}. This implies that, among other challenges, the problem of automated candidate selection must be decisively solved.

\begin{figure*}
	\includegraphics[width=0.66\textwidth,height=0.66\textheight,keepaspectratio]{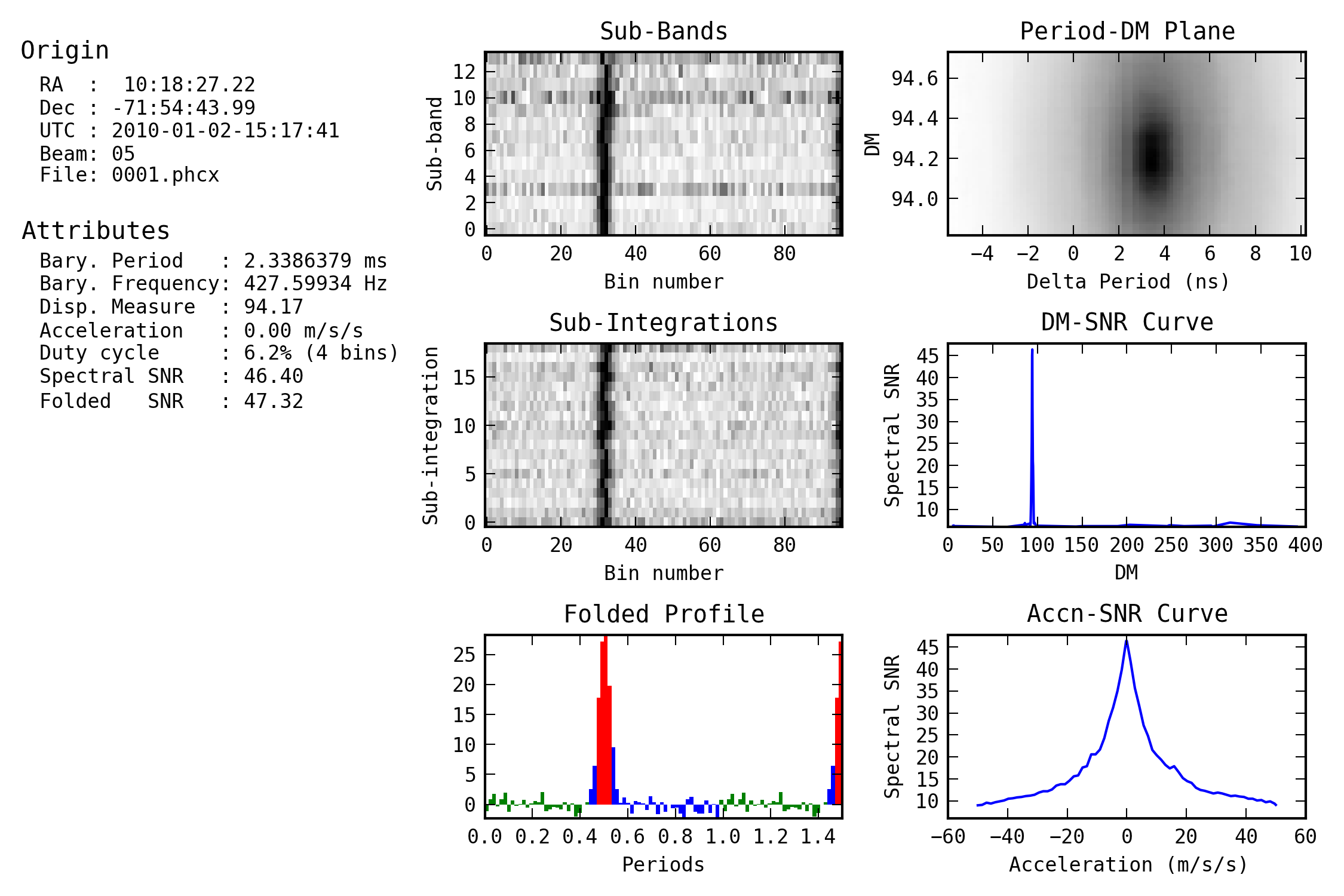}
	\centering
	\caption{Candidate information for the known pulsar J1017$-$7156, showing a number of physical diagnostics and six graphical representations describing the periodic signal it emits. See section \ref{subsec:VisualInspection} for an explanation of the graphs.}
	\label{fig:DiagnosticPlot}
\end{figure*}

	Supervised Machine Learning (ML) classifiers offer great promise in this area, and were first introduced into the field by \citet{Eatough2010}. They are general purpose methods that can be used to classify instances of multi-class data, by operating on a well-chosen set of their numerical properties called \textit{features}. They first build an internal model of the underlying statistical distributions of these features for each data class through the process of \textit{training}. In the context of supervised learning, this requires a labelled data set carefully prepared by a human expert, or \textit{training set}. Once learned, that internal representation enables the classifier to subsequently label previously unseen data. Supervised ML algorithms are well suited to classification problems where no reliable and simple rules are available to perform the task. In this work we use such a class of algorithms, namely Artificial Neural Networks (ANN), and attempt to exceed the performance of previous automated candidate selection tools, aiming in particular to correctly label pulsars with a 100\% success rate. To achieve that goal, we used a larger training data set that included 1196 pulsar observations from 542 distinct pulsars, wrote a custom ANN implementation for increased control over its training process and designed new features to describe the nature of a pulsar candidate.

	This paper first outlines the problem of candidate selection by visual inspection, and offers a review of existing methods to either reduce the workload or automate the process. In Section 3, we present a detailed introduction to ANN. Section 4 details the features we use to capture candidate characteristics, and the rationale for their design. In Section 5, our ANN implementation is evaluated on a set of candidates from the intermediate Galactic latitude area of the HTRU survey (HTRU-medlat). Its efficiency when used on new data is shown in Section 6. We conclude with a discussion of the reasons of SPINN's success, its current limitations, possible future improvements, and how future pulsar surveys should be run with ML classifiers in mind, if human intervention in classification is to be eventually reduced to a bare minimum.

\section{Pulsar candidate selection}

	Identifying new pulsar signals in observational radio data can be done either via single pulse searches \citep{McLaughlin2003}, or periodicity searches which we briefly summarize here. The first computational step, the so-called de-dispersion or DM Search, consists in correcting for the dispersive properties of the interstellar medium, which induce a delay in the observed arrival time of pulses that is both dependent on the observational frequency and the a priori unknown free-electron density integrated along the line of sight, a parameter called \textit{dispersion measure} (DM). Discovering pulsars in binary systems may also require the application of methods to compensate for the effect of orbital motion of the radio source, as its change of velocity along the line of sight causes its apparent pulse period to vary over the course of an observation, as a result of the Doppler effect. One of these methods is time domain resampling \citep{JK1991}, also referred to as Acceleration Search, working under the assumption that orbital motion during an observation sufficiently shorter than the orbital period of the source is well approximated by a constant acceleration.

	A thorough processing of the radio data therefore involves a grid search in both DM and acceleration, and for each \textit{trial} [postulated (DM, Acceleration) pair], the time series is transformed accordingly, and periodic signals are identified using a Fast Fourier Transform (FFT). Finally, the transformed time series can be ``folded" modulo the period of every significant periodic signal found, coherently stacking and summing the train of pulses of a potential pulsar. The folding process returns the final product of a pulsar survey: candidates with their set of diagnostic information, described shortly after. More detailed information about modern pulsar searching methods can be found in the standard references \citep{LK2005,LGS2006}.

\subsection{Visual inspection of candidates}
\label{subsec:VisualInspection}

	Fig.\ref{fig:DiagnosticPlot} shows the diagnostic information for a known pulsar that exhibits all the typical characteristics. The plots in the left-hand column describe, from top to bottom, the pulse in different bands of observed frequencies (sub-bands plot), the evolution of the pulse during the observation (sub-integrations plot), and the ``folded profile" which is the pulse averaged across all the observed frequencies for the entire observation. A pulsar is expected to emit in a broad range of wavelengths, with its signal remaining visible for most of the observation with a stable pulse shape. Most pulsars also display a folded profile made of a single narrow peak, although wide and/or multi-peak profiles are not uncommon. The right-hand column of plots contains from top to bottom: the Period-DM plane, which represents the evolution of the signal-to-noise ratio (S/N) of the signal as it is folded with slightly different values of period and DM. Darker colors denote a brighter signal. Below, the DM-S/N and Acceleration-S/N curves summarize the results of earlier DM and Acceleration trials, before the time series was folded, associating the S/N of the candidate in the Fourier domain with DM and Acceleration trial values. These plots are used to determine that the signal of a prospective pulsar is associated with well-defined and unique values of acceleration, period and DM. For the latter, an unambiguously non-zero value strongly indicates an extra-terrestrial origin for the source.

\subsection{Existing automated methods}

	Visual inspection of every candidate produced by a modern survey is no longer a reasonable option. As an example, our latest processing of the HTRU intermediate latitude survey \citep{Barr2014} returned 4.3 million candidates, which, at a very optimistic rate of one candidate per second, would require approximately 1200 person hours to classify. That proposition can be made even less economically interesting in the case of re-processing of data analyzed one or more times before, in which the number of expected new discoveries is much lower. The repetitive nature of the work also leads to errors during long inspection sessions. As a consequence, techniques to reduce the required amount of human intervention have been used for more than a decade.

	Graphical selection tools such as \textsc{reaper} \citep{Faulkner2004} and \textsc{jreaper} \citep{Keith2009} enable the user to project up to several thousand candidates at once in scatter plots, representing one of their features versus another, leading to rejection en masse of candidates not exhibiting the desired properties, for example excessively faint candidates, or ones found too close to narrow frequency bands polluted by RFI. Scoring algorithms such as \textsc{peace} \citep{Lee2013} have also been developed, combining 6 numerical candidate quality factors into one formula that produces a subjective ranking where pulsars are expected to be found close to the top. Machine learning (ML) solutions have also been proposed, first by \citet{Eatough2010}, who used an artificial neural network to classify outputs from the PMPS survey, operating on 8 to 12 numerical features extracted from candidate diagnostic information. \citet{Bates2012} applied the same technique on the HTRU survey, extending the number of features to 22. More recently, \citet{Zhu2013} combined a variety of ML algorithms that perform pattern recognition directly on candidate plots such as those shown in Fig. \ref{fig:DiagnosticPlot}, instead of first attempting to reduce them into features. In this work, we use an approach most similar to \citet{Eatough2010}.

\section{Artificial neural networks for binary classification}

\subsection{Supervised learning}

	Supervised learning is the task of inferring a real-valued function from a set of labelled data points called \textit{training examples}. A training example consists of a pair $(\boldsymbol{x}, y)$ where $\boldsymbol{x} \in \mathbb{R}^n$ is the input or feature vector, and $y$ the \textit{target value} or desired output (typically $y \in \mathbb{R}$) chosen by human experts or gathered from experimental measurements. Supervised learning can be used to solve regression problems, where the goal is to predict a continuous variable from a set of inputs, and classification problems, where one tries to assign a discrete class label to new unlabelled data points. In the context of binary classification, the two possible class labels are encoded in the target value, which may be set for example to 1 for members of the ``positive" class, and to 0 for members of the ``negative" class.

A wide range of supervised learning algorithms is available, including artificial neural networks. One of the valuable features of ANNs that motivated us to choose them is that they naturally produce a real-valued continuous output, the \textit{activation value}. While it can be easily converted to a binary class label by applying a threshold, the activation value also represents a level of confidence in the class label obtained. In the context of pulsar candidate classification, this can be used as a way to prioritize inspection and confirmation of candidates as we will see later. In this section we only present an introduction to ANNs geared more specifically towards their use as binary classifiers. For a more advanced and general overview see e.g. \citet{Bishop1995} or \citet{Rojas1996}. 

	\begin{figure}
	\includegraphics[width=0.45\textwidth,height=0.45\textheight,keepaspectratio]{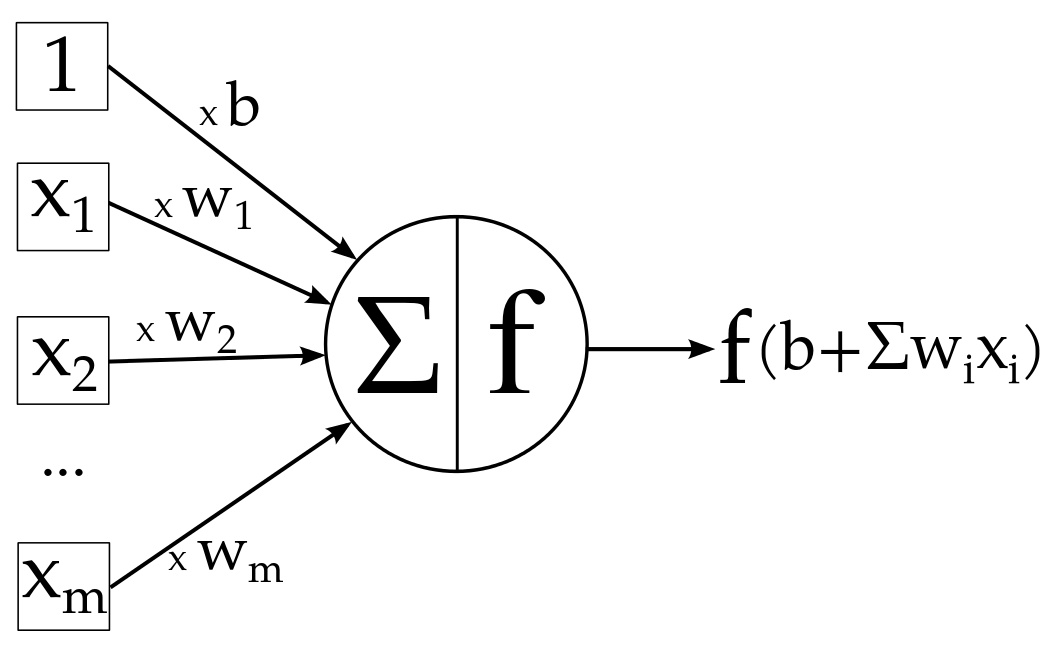}
	\caption{General model of an artificial neuron with $m$ inputs. The bias term $b$ can be seen as a weight operating on an extra constant input equal to one. The activation function $f$ usually chosen is the logistic sigmoid or similar.}
	\label{fig:ArtificialNeuron}
	\end{figure}

\subsection{Mathematical model}
	An artificial neuron is a computational model (see Fig. \ref{fig:ArtificialNeuron}) inspired by its biological counterpart, which constitutes the basic building block of a network. It is parametrized by a vector of \textit{weights} $\boldsymbol{w}$ of pre-determined dimension, a scalar \textit{bias term} $b$, and an \textit{activation function} $f$. For a given feature vector $\boldsymbol{x}$, it outputs an \textit{activation value} $a$ given by
	\begin{equation}
	\label{eq:SingleNeuron}
	a = f(\boldsymbol{w} \cdot \boldsymbol{x} + b).
	\end{equation}

	\begin{figure}
	\includegraphics[width=0.5\textwidth,height=0.5\textheight,keepaspectratio]{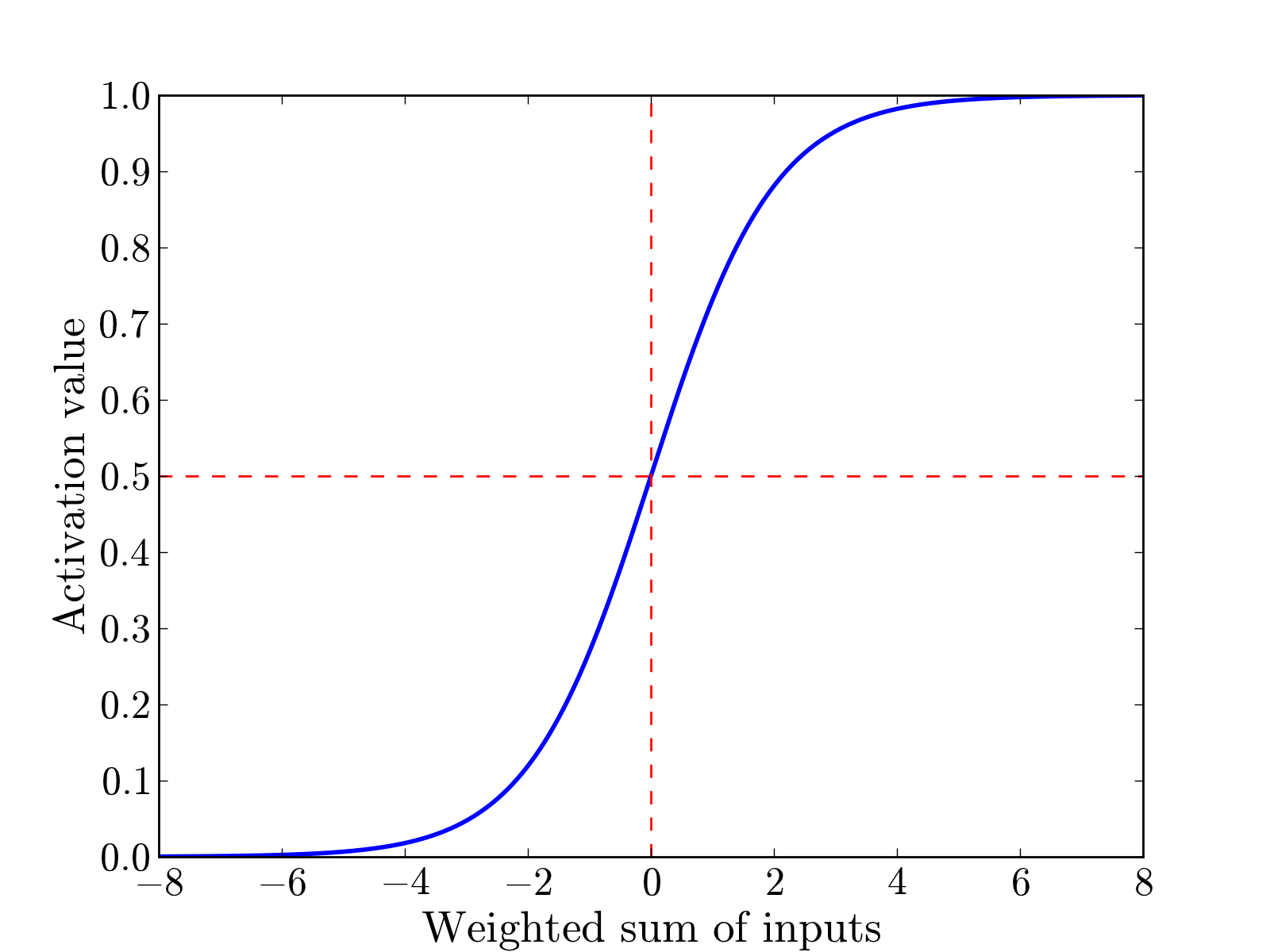}
	\caption{The logistic sigmoid activation function.}
	\label{fig:SigmoidFunction}
	\end{figure}

	Common choices for the activation function are sigmoid shaped non-linear functions such as the hyperbolic tangent or the logistic sigmoid function, the expression of the latter being
	\begin{equation}
	\label{eq:SigmoidFunction}
	f(z) = \frac{1}{1 + \exp{(-z)}},
	\end{equation}
	which takes values between 0 and 1 (see Fig. \ref{fig:SigmoidFunction}).

\begin{figure}
	\includegraphics[width=0.5\textwidth,height=0.5\textheight,keepaspectratio]{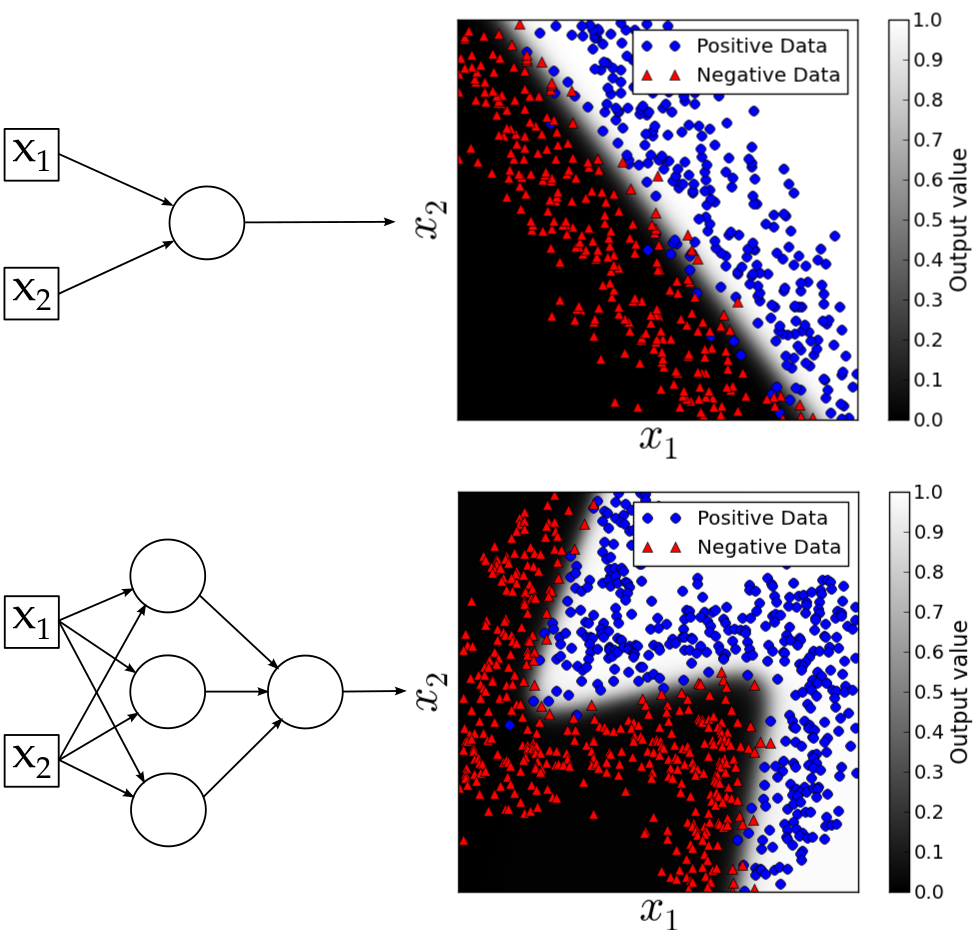}
	\centering
	\caption{Illustration of the operation of neural networks on 2D toy data sets. Top panel: a single neuron defines a splitting hyperplane in feature space, the harshness of the split increases with the norm of the weights vector. Bottom panel: combining neurons into layered networks allows complex decision boundaries to be carved. Here the three linear separations defined by the first layer of neurons are readily visible. Bias terms are not represented in network diagrams for clarity.}
	\label{fig:NeuronHyperplanes}
	\end{figure}

	A very useful geometric interpretation is to visualize a single artificial neuron as defining a separating hyperplane in feature space (Fig. \ref{fig:NeuronHyperplanes}, top panel), with a normal vector $\boldsymbol{w}$ defining its orientation, and the bias term $b$ defining its altitude at the origin. Note that the norm of the weights vector $\|{\boldsymbol{w}}\|$ is a meaningful parameter on its own, despite not having any effect on the orientation of the hyperplane: it defines its ``sharpness".

	 Unsurprisingly, an individual neuron performs poorly on non linearly separable data, but any number of them can be connected into layered networks capable of carving boundaries of arbitrarily high complexity in feature space (Fig. \ref{fig:NeuronHyperplanes}, bottom panel)

\subsection{Training}
	Training is the process of finding an adequate set of weights for the given classification problem. This is posed as an optimization problem, where a \textit{cost function} or \textit{loss function}, which measures the discrepancy between target values and actual outputs of the network on the training set, must be minimized with respect to the weights and biases of the whole network. A common choice of cost function is the mean squared error

	\begin{equation}
	\label{eq:CostFunction}
	E(w_{ij}^{(l)}, b_{j}^{(l)}) = \frac{1}{m} \sum_{k=1}^m (a_{k} - y_{k})^2,
	\end{equation}

	where $w_{ij}^{(l)}$ and $b_{j}^{(l)}$ are respectively the weights and bias term of the $j$-th neuron in layer number $l$, $y_{k}$ and $a_{k}$ are respectively the target value and activation value for training example number $k$, and $m$ the total number of training examples. The cost function is minimized using gradient descent, starting from a random initialization and going through iterations where the following three steps are performed in succession

	\begin{enumerate}
	\item Compute the activation values of the network on the data set, and the differences with the target values.

	\item Compute the derivative of the cost function with respect to the network parameters (weights and biases), using the \textit{backpropagation algorithm} (see below).

	\item Correct every network parameter $\xi$ with the following update rule, where $\eta$ is the \textit{learning rate}:
		\begin{equation}
		\label{eq:GradientDescent}
		\xi := \xi - \eta \frac{\partial E}{\partial \xi}.
		\end{equation}
	\end{enumerate}

	The backpropagation algorithm \citep[see e.g.][for a description and proof]{Rojas1996} is a very computationally efficient way to compute the gradient of the cost function with respect to the weights and biases of the network, that historically made the training of large networks tractable.

\subsection{Regularization}
	The training process only yields the best network weights and biases to properly label the training set, which does not necessarily imply optimal classification performance on unseen data. The model learned may capture not only legitimate patterns in the data, but also fit irregularities specific to the training set (due for example to its limited size), a situation referred to as \textit{overfitting}. Regularization consists in limiting the complexity of a model to improve its ability to generalize to new data. One such method that we used is \textit{L2 weight decay}, where a penalty term is introduced into the neural network cost function \eqref{eq:CostFunction} usually written as

	\begin{equation}
	\label{eq:CostFunctionPenalized}
	E(w_{ij}^{(l)}, b_{j}^{(l)}) = \frac{1}{m} \sum_{k=1}^m (a_{k} - y_{k})^2 + \lambda \sum{w_{ij}^{(l) 2}},
	\end{equation}
	with $\lambda$ being the \textit{weight decay} parameter. The net effect is to prevent the weights of the network from growing excessively large during training, therefore simplifying the decision boundary shape in feature space. The optimal value of $\lambda$, along with the optimal number of neurons to use, is found through grid search and cross-validation, described in Section 5.

\begin{figure*}
	\includegraphics[width=\textwidth,height=\textheight,keepaspectratio]{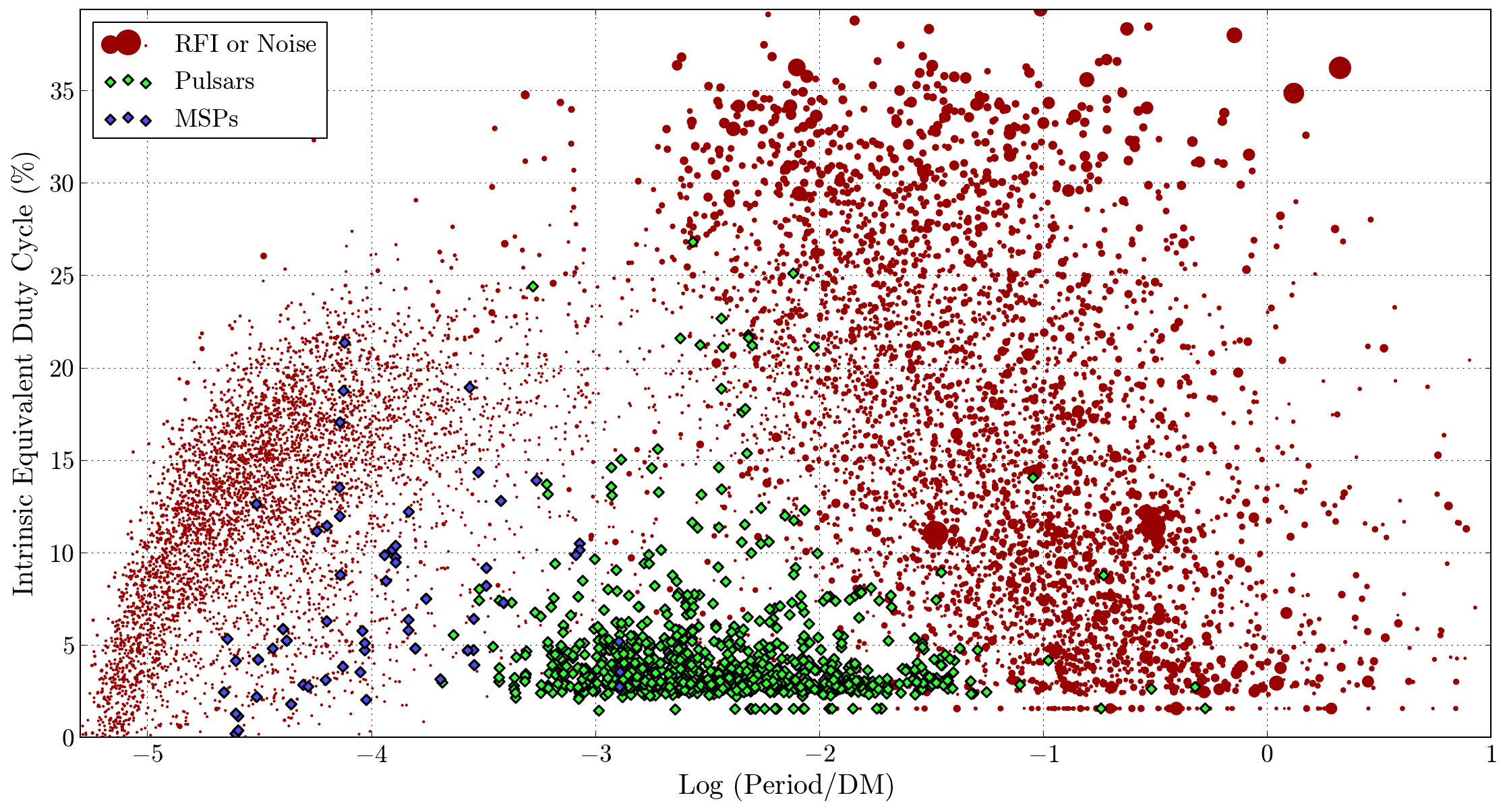}
	\centering
	\caption{HTRU Medlat training data represented in the plane $\log(\mathrm{P/DM})$ and intrinsic equivalent duty cycle, showing three main clusters. Millisecond pulsars (MSPs) are defined as having periods shorter than 50 ms for the purpose of this plot. RFI and noise candidates are represented with a size proportional to their signal-to-noise ratio. For readability, RFI and noise candidates with S/N lower than 7 or negative intrinsic duty cycles are not shown. This shows the existence of a good pulsar selection criterion independent from S/N, which is very valuable, and learnable by any ML algorithm.}
	\label{fig:IEWvsPDM}
\end{figure*}

\section{Feature design}
\label{sec:Features}

\subsection{Design choices}

	Features are the properties of an unlabelled data instance upon which a ML algorithm decides to which class it is most likely to belong. The main part of the present work consisted in reducing pulsar candidates into maximally relevant features, i.e. that take values as different as possible for pulsars and non-pulsars, and ensure that these features capture a wide range of information and domain knowledge of a human classifier. To ensure maximum classification performance, particularly with respect to the identification of faint pulsars, we obeyed the following set of guidelines:

\begin{enumerate}
\renewcommand{\theenumi}{(\arabic{enumi})}
	\item Reduce selection effects against faint or more exotic pulsars, especially MSPs or the ones with large duty cycles, which have been the most difficult to identify in the past \citep{Eatough2010, Bates2012}. As an example, the number of DM trials or Acceleration trials above a certain signal-to-noise threshold were found to introduce a strong and unjustified bias against short period candidates, regardless of their brightness. That feature was used in the past \citep{Eatough2010, Bates2012} but not in our own work.
	\item Ensure complete robustness to noisy data. As a result, no curve fitting to folded profiles or DM search graphs was attempted, as the results are very difficult to exploit properly in the low S/N regime, in which we are most interested.
	\item Limit the number of features to a set of very relevant ones, as the very limited sample of pulsar observations is unlikely to sufficiently cover all the degrees of freedom of a large feature space. An excessive number of features induces a reduction in classification performance. This is a facet of the ``curse of dimensionality" problem in Machine Learning known as the \textit{Hughes effect} \citep{Hughes1968}. This also implies avoiding the use of correlated features, as any extra feature must capture additional information.
	\end{enumerate}

\subsection{Features used}

\begin{enumerate}
\renewcommand{\theenumi}{(\arabic{enumi})}
	\item \textbf{Signal-to-Noise ratio of the folded profile (log-scale).}
	S/N is a measure of signal significance, which can be defined in various ways. We use the definition \citep[see e.g.][]{LK2005} given by equation \eqref{eq:SNR}. For a given contiguous pulse window $W$,
	\begin{equation}
	\label{eq:SNR}
	S/N = \frac{1}{\sigma\sqrt{w}} \sum_{p_i\in W} (p_i - \overline{b}),
	\end{equation}
	where $p_i$ is the amplitude of the $i$-th bin of the folded profile, $w$ is the width of the pulse region $W$ measured in bins, $\overline{b}$ and $\sigma$ are respectively the mean value and the standard deviation of the folded profile in the off-pulse region. The position and width of the pulse are determined by an exhaustive search that maximizes S/N. Once determined, the indices of the bins corresponding to pulse and baseline regions are retained in memory for further data processing. We also compute the equivalent width of the profile $w_{\rm eq}$ for further processing, defined by
	\begin{equation}
	\label{eq:equiv_width}
	w_{\rm eq}\max_{p_i \in W} p_i = \sum_{p_i \in W} (p_i - \overline{b}),
	\end{equation}
	which, in other words, would be the width of a top-hat pulse window that would have the same area and peak height as the original pulse.

	Since the values of S/N can span a wide range across candidates, we use its logarithm as a feature. Past a certain level, an increase of S/N does not make a candidate any more significant to a human expert on its own. All other things equal, two candidates with extremely	significant signal-to-noise ratios of 50 and 500 can be considered equally likely to be legitimate pulsars.

	\begin{figure}
		\includegraphics[width=0.5\textwidth,height=0.5\textheight,keepaspectratio]{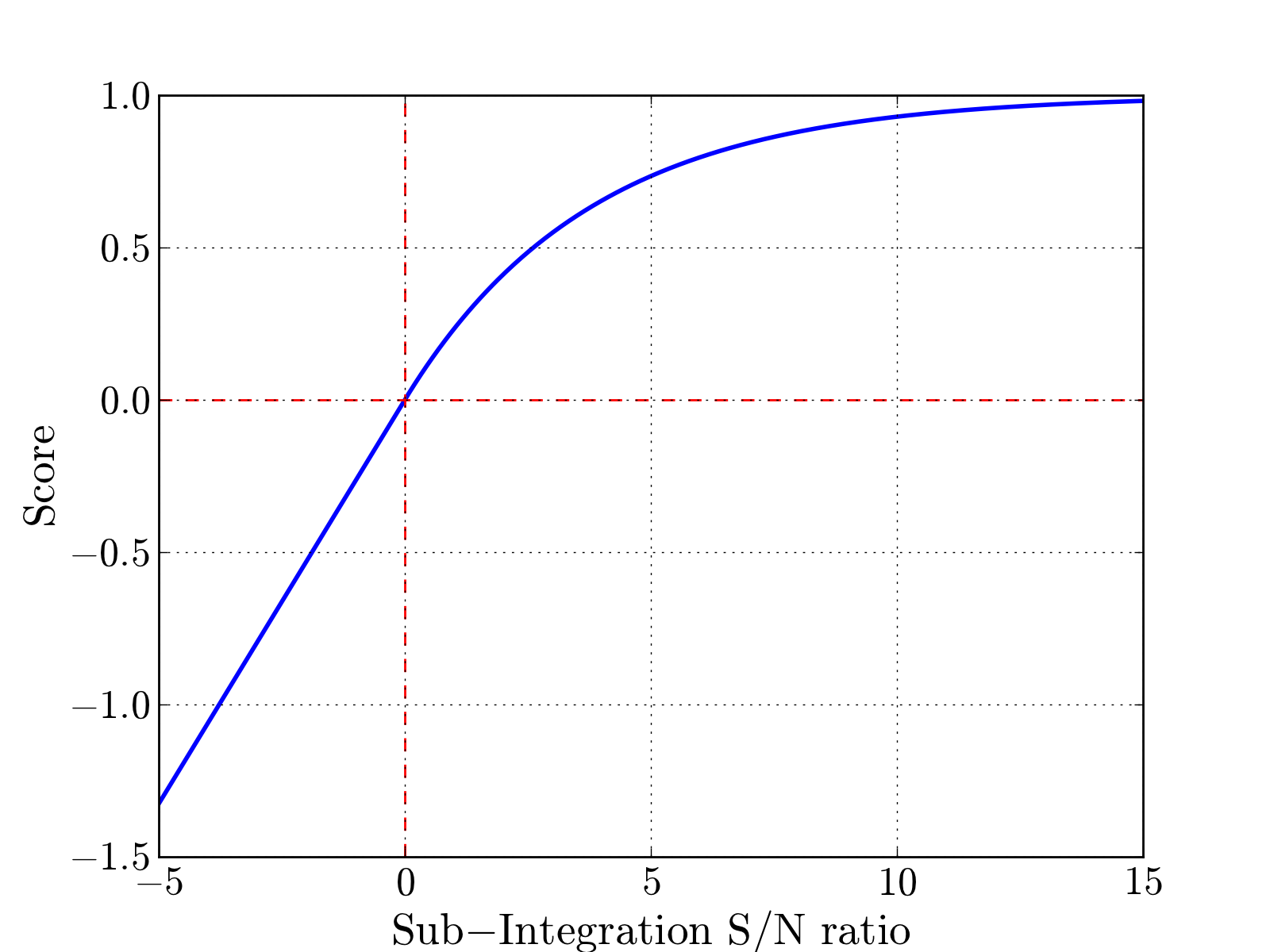}
		\centering
		\caption{An instance of the sub-integrations scoring function. Over an entire observation, persistent signals, even faint, can score higher than bright and impulsive ones.}
		\label{fig:ScoringFunction}
	\end{figure}

	\item \textbf{Intrinsic equivalent duty cycle of the pulse profile.}
	The duty cycle of a pulsar is the ratio of its pulse width $w$ expressed in seconds to its spin period. Most folded profiles of pulsars show a narrow pulse with a duty cycle typically below 5\%, while a significant amount of terrestrial signals reach much higher values up to 50\%. This can often be due to the significant amount of phase drift exhibited by artificial sources during an observation, leading to apparent smearing of their folded profiles. That being said, some pulsars, especially among the millisecond population, can have legitimately large duty cycles, which can be further increased by dispersive smearing. To avoid penalizing such objects, and further increase the usefulness of the duty cycle feature for classification, we remove the effect of dispersive smearing by defining the \textit{intrinsic equivalent duty cycle} of a candidate as
	\begin{equation}
	\label{eq:IEDC}
	D_{\rm eq} = \frac{w_{\rm eq} - \Delta\tau}{P},
	\end{equation}
	where $P$ is the period of the candidate, $w_{\rm eq}$ its equivalent width defined in \eqref{eq:equiv_width} expressed in units of time, and $\Delta\tau$ the dispersive smearing time across a frequency channel. A first-order approximation of $\Delta\tau$ is given by
	\begin{equation}
	\label{eq:smearing_time}
	\Delta\tau = 8.3\mu s (\frac{\Delta f}{\mathrm{MHz}}) (\frac{f_{\rm c}}{\mathrm{GHz}})^{-3} (\frac{\mathrm{DM}}{\mathrm{cm}^{-3}\mathrm{pc}}),
	\end{equation}
	where $\Delta f$ is the width of an observational frequency channel, $f_{\rm c}$ the centre observation frequency, and DM the dispersion measure of the candidate. Negative intrinsic equivalent duty cycle values are possible, if the dispersive smearing time exceeds the equivalent width. Strongly negative values of $D_{\rm eq}$ are not expected for a genuine astronomical signal, and this constitutes an extra selection pattern that can be learned by a ML algorithm.

	\item \textbf{Ratio between barycentric period and dispersion measure (log-scale).}
	As far as the HTRU data is concerned, the most pulsar-like RFI candidates (bright and persistent in time) 
	tend to appear at periods longer than 1 second,	and at dispersion measures close to zero. 
	A way to combine these two selection criteria into one is by considering the ratio between period and DM.
	As we did for S/N, since the values of this ratio span a large range across the pulsar and RFI population, we actually use the logarithm of the ratio 
	between period and DM as a feature. As shown in Fig.\ref{fig:IEWvsPDM}, the combination of $\log(P/{\rm DM})$ and intrinsic equivalent duty cycle offers a powerful
	selection tool that is \textit{fully independent from S/N}, splitting clearly the data into three distinct clusters: pulsars, noise candidates (faint), and RFI (bright).
	Note that the usefulness of this feature is dependent on the RFI landscape at the place and even time of observation, and its portability to other surveys is unknown.

	\item \textbf{Validity of optimized dispersion measure.}
	Pulsars can have a wide variety of dispersion measures while their RFI counterparts usually exhibit DM values very close to zero, but no other truly selective pattern
	based solely on DM can be found. Therefore we define the \textit{validity of dispersion measure} as
	\begin{equation}
	V_{\rm DM} = \tanh({\rm DM} - {\rm DM}_{\rm min}).
	\end{equation}
	The purpose of this feature is to ensure that the classifier learns to very strongly reject candidates with a DM below a certain threshold,	below which no pulsars are ever found. We used $\mathrm{DM}_{\rm min} = 2$ for HTRU-medlat data.

	\item \textbf{Persistence of signal through the time domain.}
	A genuine pulsar is expected to be consistently visible during most of an observation, and this provides a selection criterion against impulsive man-made signals. Refining an interesting idea proposed by \citet{Lee2013}, we attribute a ``score" to every sub-integration of the candidate, based on its S/N (see equation \ref{eq:SNR}) measured with respect to the pulse window and baseline region defined by the folded profile. Note that negative S/N values are possible if signal is found outside the expected window, a common property of RFI. The scoring function (see Fig. \ref{fig:ScoringFunction}) is defined as
	\begin{equation}
	\chi(s) = 
	\begin{dcases}
	    1 - \exp(-\frac{s}{b}) & \text{if } s\geq 0 \\
	    \frac{s}{b}            & \text{otherwise}   \\
	\end{dcases}
	\label{eq:subints_scoring}
	\end{equation}
	where $s$ is the signal-to-noise ratio of the candidate in a sub-integration, measured as described above, and $b$ the benchmark signal-to-noise ratio which is a user-defined parameter. The average of the scores obtained through all sub-integrations constitutes the persistence of the candidate through the time domain. Note that $b$ should be chosen low enough to filter out signals visible only for a small fraction of the observation, but not so much as to excessively penalize the class of ``nulling" pulsars \citep{Backer1970} that can become invisible for a part of the observation. We found a sensible, albeit arbitrary choice to be

	\begin{equation}
	b = \frac{2 S_{\rm min}}{\sqrt{n_{\rm sub}}},
	\label{eq:benchmark_snr}
	\end{equation}
	with $S_{\rm min}$ being an estimation of the overall signal-to-noise ratio of the faintest pulsars still clearly visible to the trained eye, and $n_{\rm sub}$ the total number of sub-integrations. We set $S_{\rm min} = 8$, also in accordance with the fact that no pulsar discovered with a S/N below 9.5 was ever confirmed over the course of the HTRU-medlat survey.

	\item \textbf{Root-mean-square distance between the folded profile and the sub-integrations.}
	The ``persistence through time" feature is insufficient on its own to capture the information that RFI signals tend to show some amount of drift in phase or even shape changes during an observation, enough to easily betray their non-astronomical nature even to a moderately well-trained human eye. To alleviate this problem, 	we define a measure of the variability of the pulse shape though the observation. 
	To compute it, we first normalize the folded profile to values between 0 and 1, and also normalize individually every sub-integration in the same fashion.
 	Let $p_i$ be the value of the $i$-th bin of the folded pulse profile, and $s_{ij}$ be the value of the $i$-th bin of the $j$-th sub-integration.
	Let $W$ be once more the set of bin indexes that constitute the pulse window in the folded profile, and $w$ the pulse window width.
	We then simply define the root mean square distance between the folded profile of the candidate and each of its sub-integrations as:
	\begin{equation}
	D_{\rm RMS} = \sqrt{\frac{1}{w n_{\rm sub}} \sum_{i \in W} \sum_{j}(p_i - s_{ij})^2 }
	\label{eq:rms_distance}
	\end{equation}
	This feature helps characterize persistent RFI in the medium to high S/N regime. 

	As a final word, a ``Persistence of signal through the frequency domain" similarly defined as its time domain counterpart was tried and initially believed to provide a very useful selection criterion to separate broadband pulsar signals from all the others. It was eventually removed from the feature set, as its addition proved slightly detrimental to classification performance on HTRU medlat data. We attributed this effect to the relative absence of candidates that arise from narrowband radio frequency interference, but it might be useful for other surveys facing different RFI populations. Computing persistence features as defined earlier relies on comparing the brightness of the signal vs. a benchmark, which is constant in the time domain. For surveys with large fractional bandwiths, correcting for the average spectral index of pulsars when computing that benchmark in the frequency domain could be beneficial.
	
\end{enumerate}

\section{Classification performance on HTRU medlat data}

\subsection{Training set}

	Training examples were gathered from the outputs of a new processing of HTRU medlat data, performed as a test run of the new high-performance GPU-based \textsc{peasoup} pipeline \citep{Barr2014}, with acceleration searching enabled. DMs from 0 to 400 $\rm{cm}^{-3}\rm{pc}$, and accelerations from $-50$ to $+50$ $\rm{m/s}^2$ were searched, yielding in excess of a thousand candidates in each of the 95,725 beams. Signal periods simultaneously found in three beams or more of the same pointing were ignored, and the resulting 50 brightest candidates in every beam were folded, with the exception that any signal with a Fourier-domain S/N in excess of 9 was automatically folded as well. This processing strategy returned 4.34 million folded candidates. Each of them was individually matched against the ATNF pulsar catalogue \citep{Manchester2005} to label all known pulsars (and their harmonics) that were found by the pipeline. A total of 1196 observations of 542 distinct known pulsars were identified after being carefully reviewed by eye to confirm their nature. They constituted the positive class of the ANN training set, to which we added 90,000 non-pulsar observations picked at random as a negative class, to obtain a varied and representative sample of the population of spurious folded candidates. We assumed that none of these were undiscovered pulsars: in past HTRU medlat processings, approximately 100 new pulsars were found in 10 million folded candidates, a discovery rate of 1:100,000. In our re-processing set of candidates, it is expected to be significantly lower.

	The training set contains pulsars with varied spin periods, duty cycles, and signal-to-noise ratios. Among them, 77 have periods shorter than 50 ms and 46 have duty cycles larger than 20\%. A total of 78 observations of pulsars have folded signal-to-noise ratios below 10, down to a minimum of 7.3. For the sake of comparison, no pulsar discovered with a folded S/N below 9.5 was ever confirmed over the entire course of the HTRU survey. Low S/N pulsars are difficult to distinguish from noise fluctuations even by eye, and the limited observation time available imposes a conservative S/N selection threshold (a consensual value is 10) on the folded candidates to confirm. Despite these limitations, fainter pulsar observations were kept in the training set to ensure maximum sensitivity.

\begin{figure*}
	\includegraphics[width=\textwidth,height=\textheight,keepaspectratio]{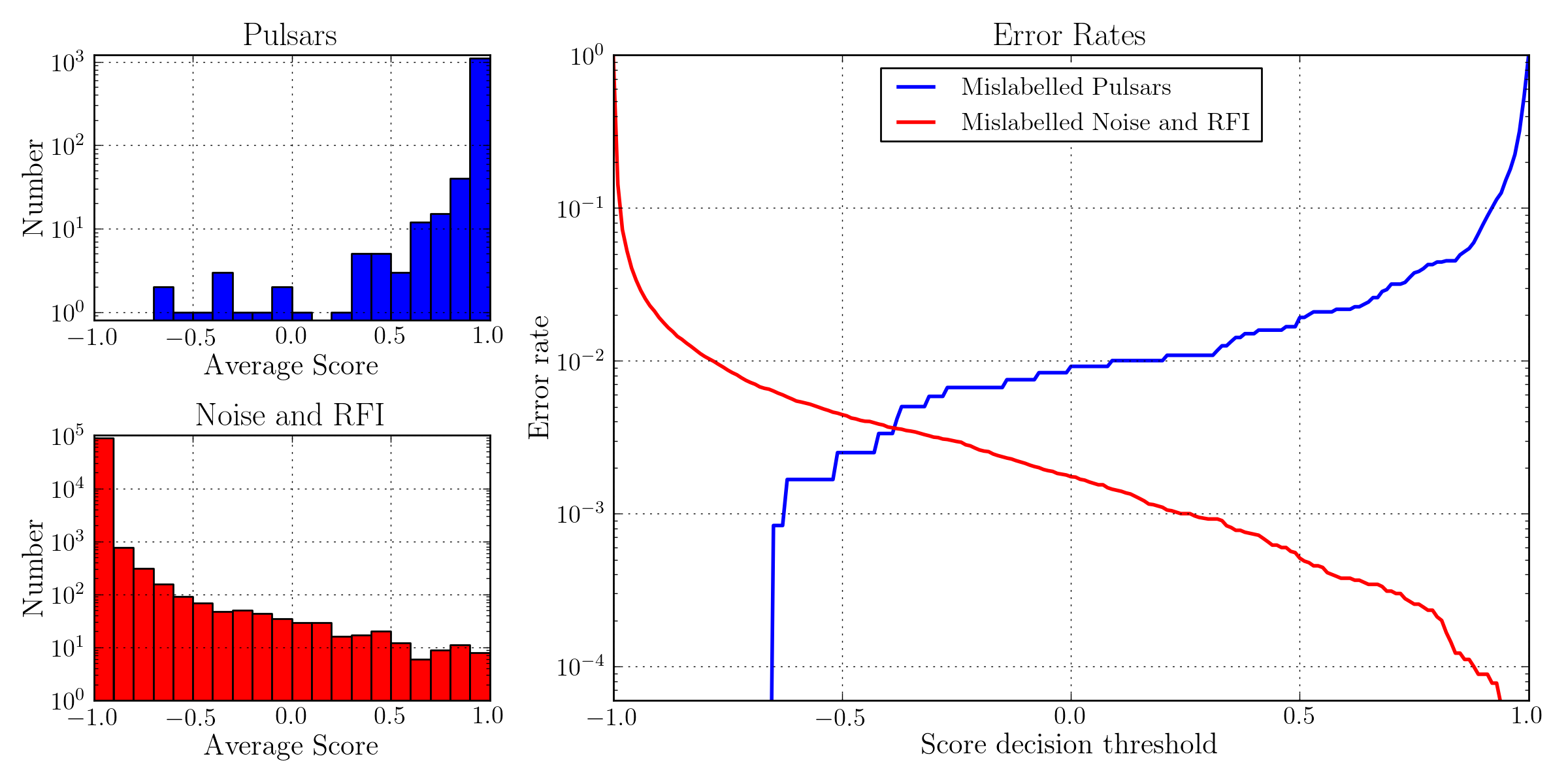}
	\centering
	\caption{Results of 20 iterations of 5-fold cross-validation. Every time a candidate was in the test set, and therefore \textit{not} trained upon, its ANN activation value, that we also refer to as score, was recorded. Left: Histograms of average ANN scores obtained by every candidate from our labelled data. Right: Expected error rates on new data as a function of the score decision threshold chosen, whereby candidates scoring above that threshold are classified as pulsars.}
	\label{fig:XVResults}
\end{figure*}

\subsection{Implementation details}

	To obtain fine control over the training process, a custom ANN implementation was written. We followed some practical recommendations detailed at length in \citet{LeCun1998}, which we enumerate here.
	\begin{enumerate}
	\renewcommand{\theenumi}{(\arabic{enumi})}
	\item Feature scaling was performed before training, ensuring that every individual feature has zero mean and unit standard deviation over the entire training set.
	\item The hyperbolic tangent activation function was used instead of the logistic sigmoid, yielding activation values ranging from $-1$ to $+1$.
	\item The ANN was trained using ``mini-batches", whereby during each training epoch the network is presented only with a small, changing subset of the training set. This is not only a much faster process than standard ``batch" training, but also its noisy nature can help to avoid local minima of the network cost function.
	\end{enumerate}
	To overcome the large class imbalance of the training set, we oversampled the pulsars to obtain a 4:1 ratio of non-pulsars to pulsars, so that pulsar candidates were ``seen" much more often by the ANN during training, while preserving the variety of non-pulsars.

\subsection{Choice of performance metrics}

	Obviously, the goal of an automated classifier is to identify the largest possible fraction of pulsars while returning a minimal amount of mislabelled noise or RFI (false positives). The two natural performance metrics for this problem, measured on a test sample of labelled data, are therefore:

	\begin{equation}
	\textrm{Recall} = \frac{\textrm{True Positives}}{\textrm{True Positives} + \textrm{False Negatives}}
	\label{eq:recall}
	\end{equation}

	\begin{equation}
	\textrm{False positive rate} = \frac{\textrm{False Positives}}{\textrm{True Negatives} + \textrm{False Positives}}
	\label{eq:fposrate}
	\end{equation}

	The true positives are the positive examples correctly labelled as such, and false negatives are the positive examples incorrectly labelled as negatives. Recall is therefore the fraction of positives properly labelled. Likewise, the false positive rate is the fraction of negatives mislabelled as positives. Other common metrics such as \textit{Accuracy} or \textit{F-Score} depend explictly on class imbalance (the ratio of positive to negative examples in the test data), and as such are not suitable if we are to compare classifier performances across different test samples or even different pulsar surveys. Since pulsars remain arguably rare objects up to this day, missing any of them carries a heavy cost, and emphasis must be put on maximising recall above all else. The amount of visual inspection required to select folded candidates for confirmation is proportional to the false positive rate of the classifier, for which low values are desirable.

\subsection{Cross-validation}

	We performed a 5-fold cross-validation procedure, to choose an optimal network architecture and weight decay, and evaluate classification performance. This consists in randomly partitioning all of the labelled data into 5 equally sized subsets, each of them being successively held out as a test set, while only the remaining four are used to train the ANN upon. This ensures that performance is always evaluated on data unseen during training, and that over the five iterations (``folds") of this procedure every candidate ends up in the test set exactly once, at which point its ANN activation value or ``score" (that takes continuous values between -1 and +1) was recorded. We repeated the entire process 20 times, with different random data partitions, obtaining a representative average score for every candidate from our labelled sample that does not depend on a specific training/test set split.

	The resulting list of scores allowed to determine, for every score decision threshold between $-1$ and $+1$, what were the associated Recalls and False positive rates, ie. how many pulsars were below that decision threshold, and how many RFI or noise candidates were above. When later deploying the ANN on new data, this gives an estimate of how far down the score ladder candidates should be inspected by eye, where the tradeoff between Recall and False positive rate is left to the user's discretion. Furthermore, examining consistently low scoring pulsars gave insight into which ones the ANN was biased against, which we will discuss later. Finally, by repeating this whole procedure with various network architectures and weight decay values, we were able to settle on an optimal ANN configuration. It was chosen so as to minimize the number of non-pulsars scoring better than the worst-scoring pulsar, that is minimize the false positive rate at 100\% Recall. A simple two-layered 8:1 network (8 hidden units, one output unit) was found to yield the best results, with performance progressively degrading with more units.

\subsection{Classification Performance}

	Fig. \ref{fig:XVResults} shows the distribution of scores obtained by all candidates in our labelled sample during cross-validation, and illustrates the Recall / False positive rate tradeoff. Table \ref{table:PerformanceSummary} summarizes expected classification performance for various score decision thresholds. The score distribution of pulsars shows a long tail where no more than a dozen low-scoring pulsars are responsible for the major part of the false positive rate. Their close inspection reveals that they always share at least two of the following characteristics, making them similar to noise or RFI candidates with respect to the feature space we used: large duty cycles in excess of 20\%, low S/N (below 9), high value of $\log(P/\textrm{DM})$. A few other low-scoring pulsar observations were found to exhibit abnormally low persistence through time, being rendered invisible during a part of the observation by short bursts of RFI.

\begin{table}
\caption{Classification performance during cross-validation, summarizing some key values from Fig. \ref{fig:XVResults}.}
\centering
\begin{tabular}{lrr}
	\toprule
	Recall & Score threshold & False Positive Rate\\
	\midrule
	100\% & $-0.65$ & 0.64\% \\
	99\%  & $+0.20$ & 0.11\% \\
	98\%  & $+0.52$ & 0.05\% \\
	95\%  & $+0.86$ & 0.01\% \\
	\bottomrule
\end{tabular}
\label{table:PerformanceSummary}
\end{table}

\section{Deployment on HTRU-medlat data and Discoveries}

\begin{table*}
\centering
\caption{Credible new pulsar candidates identified after visual inspection of the best 2400 returned by SPINN, after having been deployed on the entire HTRU-medlat survey (4.34 million candidates). SPINN is sensitive to short period signals with large duty cycles, and can identify interesting candidates at low S/N, below the consensual confirmation threshold of 10. Four of these candidates have already been re-observed at Parkes Observatory and confirmed as genuine pulsars. See Fig. \ref{fig:Discoveries} for their candidate plots.}
\begin{tabular}{rcrrrrr}
	Rank & Score & S/N & Period (ms) & DM ($\rm{cm}^{-3}\rm{pc}$) & Duty Cycle (\%) & Confirmed \\	
	\midrule
	9    & +0.99 & 16.3 & 2.48023    & 57.4  & 6.2  & Yes \\
	164  & +0.96 & 9.5  & 267.26763  & 104.8 & 3.1  & -  \\
	261  & +0.94 & 17.6 & 1623.72522 & 132.6 & 14.1 & Yes  \\
	755  & +0.83 & 11.1 & 1.49269    & 232.4 & 28.1 & Yes \\
	789  & +0.82 & 8.4  & 8.47148    & 18.6  & 4.7  & - \\
	801  & +0.82 & 8.2  & 593.58567  & 160.9 & 9.4  & - \\
	826  & +0.81 & 10.8 & 89.67077   & 158.0 & 20.3 & - \\
	1254 & +0.72 & 11.3 & 1220.15740 & 87.7  & 9.4  & - \\
	1287 & +0.71 & 9.0  & 406.72994  & 43.3  & 9.4  & - \\
	1388 & +0.69 & 11.0 & 527.34434  & 40.4  & 14.1 & - \\
	1482 & +0.67 & 10.6 & 568.01602  & 75.4  & 29.7 & - \\
	1779 & +0.61 & 8.3  & 20.26131   & 94.5  & 12.5 & - \\
	1926 & +0.58 & 9.6  & 7.39151    & 34.0  & 14.1 & - \\
	2367 & +0.51 & 9.9  & 4.41080    & 70.2  & 15.6 & Yes \\
	\bottomrule
\end{tabular}
\label{table:Discoveries}
\end{table*}

	\begin{figure*}
	\includegraphics[width=\textwidth,keepaspectratio]{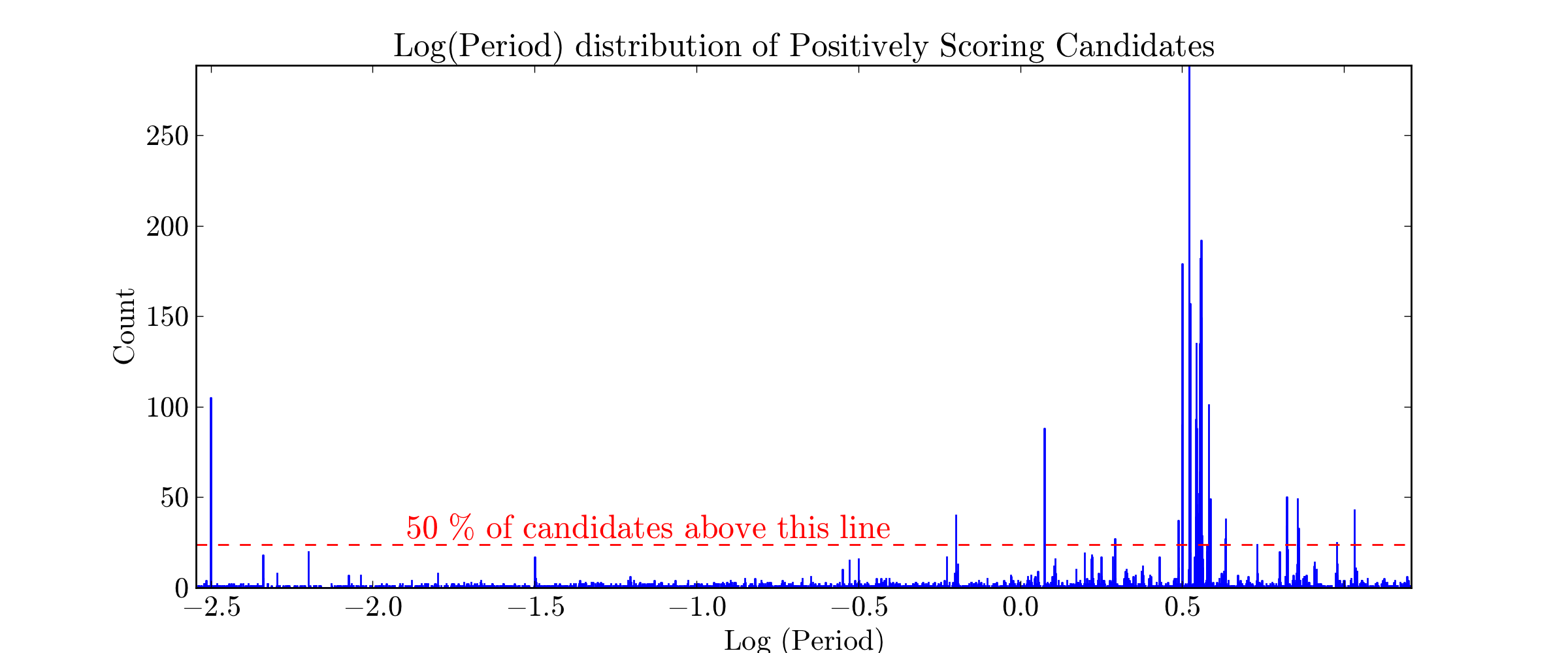}
	\centering
	\caption{Distribution in $\log(Period)$ of all previously unknown 7094 candidates that obtained a positive score during the evaluation of the entire HTRU medlat survey. Periods are expressed in seconds. One percent of the 4000 bins of the histogram presented here contain 50\% of these candidates. The fact that most pulsar-like RFI are concentrated within a very limited number of bins can be used \textit{a posteriori} to rank candidates shortlisted for inspection in a more sensible way, further increasing the discovery rate.}
	\label{fig:LogPeriodDistribution}
	\end{figure*}

	\begin{figure*}
	\includegraphics[width=0.75\textwidth,keepaspectratio]{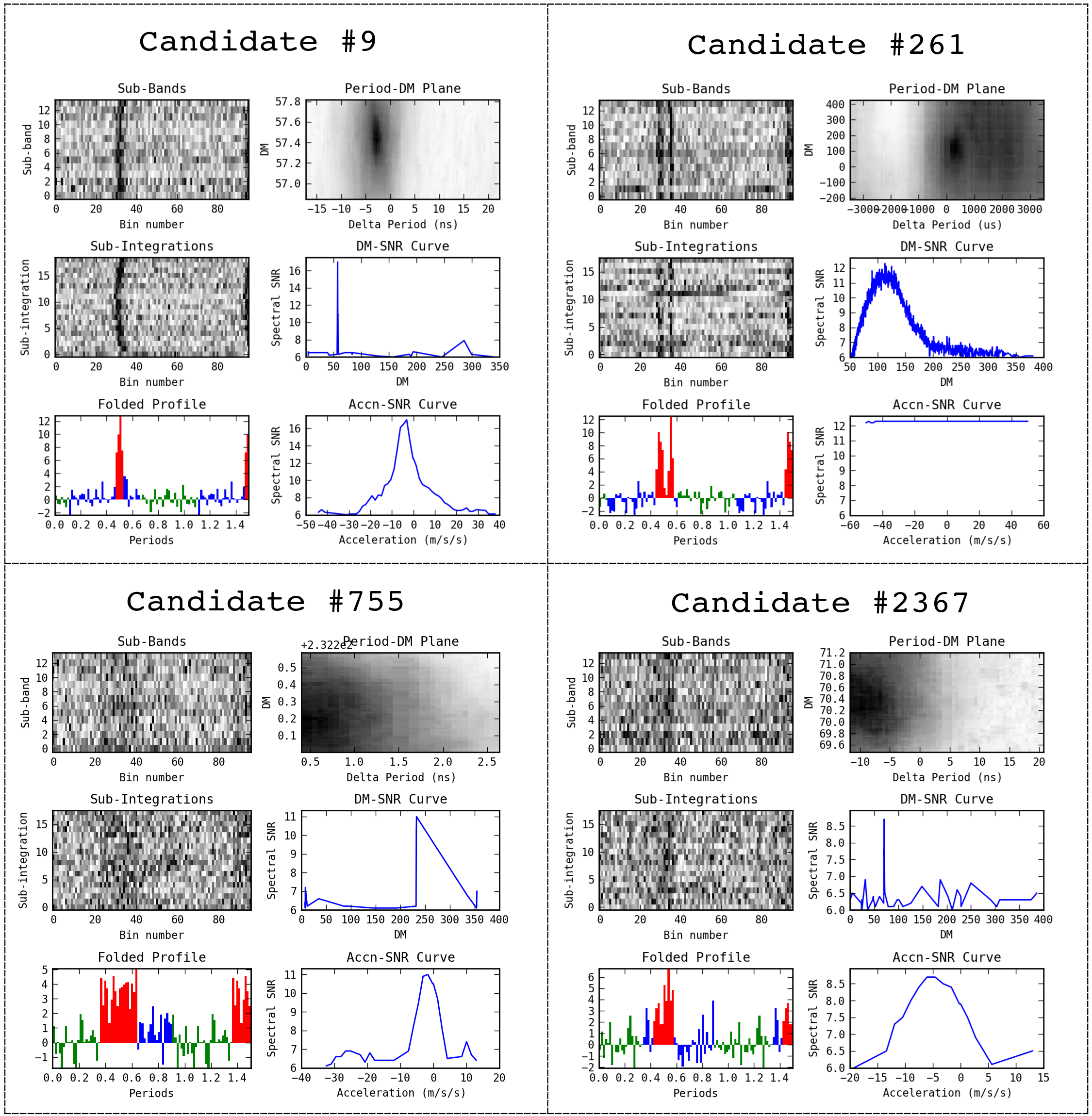}
	\centering
	\caption{Discovery plots of the four new pulsars found with the help of SPINN in HTRU-medlat observations. They were all ranked in the top 0.06\% of all 4.34 million candidates, and three of them within the top 0.02\%. These pulsars will be detailed in a future paper.}
	\label{fig:Discoveries}
	\end{figure*}

	A fully trained ANN was deployed on all 4.34 million candidates returned by the processing previously described in Section 5.1, a process that takes only 400 CPU-hours, despite being severely limited in speed by a large amount of small file I/O operations. Using 64 CPUs on Swinburne University's gSTAR cluster, this can be done overnight. Candidates were sorted by decreasing score, and known pulsars and their harmonics were removed from the list. In light of the cross-validation results (see Fig. \ref{fig:XVResults} and Table \ref{table:PerformanceSummary}), one can reasonably expect to find all potential discoveries above a score threshold of $-0.65$, which left approximately 27000 candidates to review. 

\label{subsec:VisualInspectionReport}

	So far all 2400 candidates that scored above $+0.5$ returned by SPINN have been inspected, a process that will continue as observation time to confirm possible discoveries becomes available. Table \ref{table:Discoveries} summarizes the attributes of the most promising candidates found among them. It shows that SPINN is very sensitive to pulsar-like signals down to S/N = 8, and that it can also highly rank broad pulses (duty cycles in excess of 20\%) and potential millisecond pulsars, which is a known blind spot of some previous ML solutions \citep{Eatough2010, Bates2012}. It should be noted that a considerable amount of the reviewed candidates were RFI signals with very specific periods. Fig. \ref{fig:LogPeriodDistribution} presents the distribution in $\log(\textrm{Period})$ of all previously unknown 7094 candidates that obtained a positive score. One percent of the bins account for 50\% of these candidates. One could either postpone or skip the inspection of heavily polluted period intervals, or adjust \textit{a posteriori} all ANN scores via Bayesian Inference, using the period distribution of high scoring candidates and that of known pulsars \citep{Zhu2013}. We intend to implement such a scheme in the future.

	The candidates of Table \ref{table:Discoveries} with a S/N above 9.5 have been reobserved at the Parkes Observatory and four were confirmed as new pulsars, three of which have spin periods shorter than 5 ms. Fig. \ref{fig:Discoveries} shows their candidate plots exactly as they were evaluated by SPINN. The details of these four new discoveries will be discussed in a future paper \citep{Barr2014} once their long-term coherent timing solutions have been obtained from currently ongoing observations.

\section{Discussion and Conclusion}

	We have described SPINN, an automated pulsar candidate classifier designed with maximum recall in mind. Being essentially an artificial neural network that produces a real-valued and continuous output instead of a binary class label, it can also produce a subjective ranking of candidates that can be used to prioritize visual inspection. SPINN was cross-validated on a data set containing all known pulsars found by the \textsc{peasoup} pipeline in a re-processing of HTRU-medlat and 90,000 non-pulsar candidates chosen at random. Its expected recall and false positive rates were evaluated (see Table \ref{table:PerformanceSummary}) and it was found to be capable of reducing the survey's outputs by a factor of approximately 150 while identifying all potential pulsars. Reduction factors of several thousand can be achieved at the cost of postponing a small fraction of new discoveries, an interesting prospect for future surveys.	SPINN was deployed on all candidates produced during re-processing and four new pulsars were discovered in the 2,400 candidates it ranked most highly (less than 0.06\% of the survey's output). Three of them are millisecond pulsars, one of which was found with a signal-to-noise ratio below 10. 

\subsection{The need for public training and test data for unbiased performance comparisons}

	While SPINN seems to be a significant step forward compared to previous solutions, the performance of automated candidate classifiers depends very significantly on the properties of the data they are evaluated upon. The amount of bright spurious candidates to be sifted through will be affected by the RFI landscape at the observation site, and RFI mitigation techniques used during early stages of observational data processing. The available quantity and variety of known pulsar candidates to train ML algorithms upon also plays a role, and the absence of even a handful of pulsars difficult to detect even by visual inspection in test data can skew the results heavily. Fig. \ref{fig:XVResults} illustrates this fact, as the removal of ten ``well-chosen" pulsars from our training data could have unjustifiably reduced the reported false positive rate by an order of magnitude, while obviously reducing SPINN's sensitivity. Comparisons between automated solutions are therefore limited at best, unless they are made on a common data set. To address this issue, we have made publicly available the set of candidates on which SPINN was cross-validated (see appendix \ref{sec:TrainingData}). This will allow other authors of classifiers to evaluate their own solutions, and hopefully stimulate interest in the pulsar candidate classification problem, even from Machine Learning enthusiasts not necessarily acquainted with Astronomy. A summary of reported performance of existing automated candidate classifiers is provided in Table \ref{table:PerformanceComparison} with all the previous caveats in mind.

\begin{table*}
\caption{Reported performance of existing automated candidate classifiers on various test data sets. Comparisons must be undertaken cautiously as classification performance is affected by the time and location where observational data is acquired, the RFI mitigation techniques used in data processing, and most importantly the composition of test data.}
\begin{tabular}{lllllr}

	\toprule
	Classifier & Type & Recall & False Positive Rate & Test Data Origin & Comments\\
	\midrule[1pt]

	\citet{Bates2012} & ANN & 85\% & 1\% & HTRU$^1$ Medlat & \\
	\midrule

	\multirow{2}{2.4cm}{\citet{Eatough2010}} & \multirow{2}{*}{ANN} & 93\% & 1\% & \multirow{2}{*}{PMPS$^2$} & 12:2 Network \\
	 & & 92\% & 0.5\% &  & 8:2 Network\\
	\midrule

	\multirow{3}{2.4cm}{PEACE \\ \citet{Lee2013}} & \multirow{3}{2.0cm}{Scoring algorithm} & 100\% & 3.7\% & \multirow{3}{*}{GBNCC$^3$} &  \\
	 & & 95\% & 0.34\% &  & \\
	 & & 68\% & 0.17\% &  & \\
	\midrule

	\multirow{4}{2.4cm}{PICS \\ \citet{Zhu2013}} & \multirow{4}{2.0cm}{Committee of ML algorithms} & 92\% & 1\% & PALFA$^4$ & \\
	& & 100\% & 3.8\% & GBNCC$^3$ & Trained on PALFA Data \\
	& & 100\%$^\clubsuit$ & 1.1\%$^\clubsuit$ & GBNCC$^3$ & Trained on PALFA Data \\
	& & 68\% & 0.16\% & GBNCC$^3$ & Trained on PALFA Data \\
	\midrule

	\multirow{3}{*}{SPINN} & \multirow{3}{*}{ANN} & 100\% & 0.64\% & \multirow{3}{*}{HTRU$^1$ Medlat} &  \\
	      &     & 99\%  & 0.11\% &  &  \\
	      &     & 95\%  & 0.01\% &  &  \\
	\midrule

	\multicolumn{6}{l}{$^1$ \citet{Keith2010}} \\
	\multicolumn{6}{l}{$^2$ \citet{Manchester2001}} \\
	\multicolumn{6}{l}{$^3$ \citet{Lynch2013}} \\
	\multicolumn{6}{l}{$^4$ \citet{Lazarus2013}} \\
	\multicolumn{6}{l}{$^\clubsuit$ Candidates found in RFI-polluted frequency bins had their final score reduced.} \\
	\bottomrule
\end{tabular}
\label{table:PerformanceComparison}
\end{table*}

\subsection{The need for optimal feature sets and appropriately complex models}

	ML algorithms operate on numerical features that carry no label or context, and unlike human classifiers, cannot rely on any domain knowledge associated with these features. Therefore, the statistical distributions of these features for different classes (pulsars and non-pulsars) should overlap as little as possible, so that these classes are more easily separated in feature space (see Fig. \ref{fig:NeuronHyperplanes} for a visual interpretation on a toy example). Different features were tried and discarded in the context of this work and the best subset of them, reported in Section \ref{sec:Features}, selected through cross validation with maximum recall in mind.

	Also, while this may appear counter-intuitive, ML algorithms do not always perform better with more features. For a given amount of training data, larger feature spaces will be more sparsely sampled, and the data distributions more difficult to infer accurately. The decision boundaries separating these distributions must be of appropriate complexity, which, in the case of Artificial Neural Networks, increases with the number of layers and neurons in the network. We attribute part of SPINN's success to its relatively simple internal model of a pulsar candidate. The low number of features and neurons used are in accordance with the limited amount of pulsar observations available. This can be quantitatively supported by the fact that the PICS (without score adjustment) and PEACE classifiers show almost identical performance on candidates taken from GBNCC data, as shown in Table \ref{table:PerformanceComparison}. PICS makes use of a committee of ML classifiers, including deep neural networks containing in excess of eight thousand neurons, while PEACE relies on a linear function of six numerical features. In this case, the large increase in model complexity did not translate into a significant improvement of classification performance.

\subsection{Limitations}

	SPINN's limitations are closely related to the features it relies upon. The $\log(P/DM)$ feature has provided a very simple and efficient selection criterion against artificial signals on HTRU-medlat data partly because a large majority of RFI appears at periods in excess of one second (Fig. \ref{fig:LogPeriodDistribution}), a rule not guaranteed to hold true in other surveys. Cross-validation also indicated that SPINN carries a bias against pulsar signals showing a large duty cycle, and scores them even lower if they also exhibit a high $\log(P/DM)$ value, as any pulsar signal with these two properties becomes difficult to distinguish from spurious ones in SPINN's feature space (Fig. \ref{fig:IEWvsPDM}). Characterization of RFI is therefore largely incomplete, which was confirmed by the visual inspection of high scoring candidates (see section \ref{subsec:VisualInspectionReport} and Table \ref{table:Discoveries}). With the known pulsars and their harmonics ignored, the remaining shortlist was dominated by artificial signals that had pulsar-like properties with respect to SPINN's feature space, yet their nature was often recognizable to the eye. Some signals were found to drift erratically over time, or to emit only in a set of narrow frequency bands. While SPINN has been successful in finding new pulsars, its false positive rate would have to be further reduced by about two orders of magnitude for it to be a match for the human eye in terms of accuracy, and while this is an ambitious goal, it leaves the problem of automated pulsar candidate classification largely open.

\subsection{Perspectives and Future work}

	We have previously suggested that SPINN's success comes from its low-complexity model of a pulsar candidate, relying on a lower number of features and neurons than what has been previously used, a consequence of the limited number of known pulsars observable from any given site on Earth. This would have two main implications for any present or future ML solutions.

	Firstly, collecting as many pulsar observations as possible for training should be a top priority. Observing known sources multiple times could be a viable option, purposely not pointing exactly at them to simulate a blind all-sky search, or purposely processing only a part of such an observation to reduce the S/N of the output candidate. Artificially generating credible pulsar candidates is another possibility previously proposed several times \citep{Eatough2010, Bates2012, Zhu2013}. This would come with the challenges associated with realistically simulating the varying properties of pulsar signals and all forms of noise or interference that affect them. 

	Secondly, additional efforts should be undertaken to improve the quality of the folded data if we are to increasingly rely on ML for pulsar candidate selection. Dealing with candidates folded with wrong periods, dispersion measures or accelerations vastly increases the number of degrees of freedom of the classification problem by perturbating the candidate plots in various ways. While such mistakes can be easily spotted by eye with domain knowledge and the wealth of information available in candidate plots, ML algorithms would require more features to achieve the same, and enough training data to properly sample the space of possible processing errors. Instead, extra features that can be afforded would find better use in improving RFI characterization, which will be the main focus of our future efforts to further increase classification performance.

\section*{Acknowledgements}

	The Parkes Observatory is part of the Australia Telescope which is funded by the Commonwealth of Australia for operation as a National Facility managed by CSIRO. This work was supported by the Australian Research Council Centre for Excellence for All-sky Astrophysics (CAASTRO), through project number CE110001020. The re-processing and scoring of the HTRU medlat survey made extensive use of the GPU Supercomputer for Theoretical Astrophysics Research (gSTAR) hosted at Swinburne University, and funded by a grant obtained via Astronomy Australia Limited (AAL). We thank the anonymous referee for their helpful remarks.

\appendix
\section[]{Training data set}
\label{sec:TrainingData}

The entire training data set that was used to cross-validate SPINN in this work is available at \url{http://astronomy.swin.edu.au/~vmorello/}

Candidates are provided in the PulsarHunter Candidate XML format (PHCX).


\begin{thebibliography}{99}

	\bibitem[\protect\citeauthoryear{Backer}{1970}]{Backer1970} 
	Backer D.~C., 1970, Nature, 228, 42 

	\bibitem[\protect\citeauthoryear{Barr et al.}{in preparation}]{Barr2014} 
	Barr E.~D., et al., in prep.

	\bibitem[\protect\citeauthoryear{Bates et al.}{2012}]{Bates2012} 
	Bates S.~D., et al., 2012, MNRAS, 427, 1052 

	\bibitem[\protect\citeauthoryear{Bishop}{1995}]{Bishop1995} 
	Bishop C. M., 1995, Neural Networks for Pattern Recognition,	Oxford University Press

	\bibitem[\protect\citeauthoryear{Eatough et al.}{2010}]{Eatough2010} 
	Eatough R.~P., Molkenthin N., Kramer M., Noutsos A., Keith M.~J., Stappers B.~W., Lyne A.~G., 2010, MNRAS, 407, 2443 

	\bibitem[\protect\citeauthoryear{Faulkner et al.}{2004}]{Faulkner2004}
	Faulkner A.~J., et al., 2004, MNRAS, 355, 147 

	\bibitem[\protect\citeauthoryear{Hughes}{1968}]{Hughes1968}
	Hughes G., 1968, IEEE Transactions on Information Theory, 14, 55

	\bibitem[\protect\citeauthoryear{Johnston \& Kulkarni}{1991}]{JK1991} 
	Johnston H.~M., Kulkarni S.~R., 1991, ApJ, 368, 504 

	\bibitem[\protect\citeauthoryear{Keith et al.}{2009}]{Keith2009} 
	Keith M.~J., Eatough R.~P., Lyne A.~G., Kramer M., Possenti A., Camilo F., Manchester R.~N., 2009, MNRAS, 395, 837 

	\bibitem[\protect\citeauthoryear{Keith et al.}{2010}]{Keith2010} 
	Keith M.~J., et al., 2010, MNRAS, 409, 619 

	\bibitem[\protect\citeauthoryear{Lazarus}{2013}]{Lazarus2013} 
	Lazarus P., 2013, IAUS, 291, 35 

	\bibitem[\protect\citeauthoryear{LeCun et al.}{1998}]{LeCun1998} 
	Y. LeCun, L. Bottou, G. Orr and K. Muller: Efficient BackProp, in Orr, G. and Muller K. (Eds), Neural Networks: Tricks of the trade, Springer, 1998

	\bibitem[\protect\citeauthoryear{Lee et al.}{2013}]{Lee2013} 
	Lee K.~J., et al., 2013, MNRAS, 433, 688 

	\bibitem[\protect\citeauthoryear{Lorimer \& Kramer}{2005}]{LK2005} 
	Lorimer D. R., Kramer M., 2005, Handbook of Pulsar Astronomy, Cambridge University Press

	\bibitem[\protect\citeauthoryear{Lynch et al.}{2013}]{Lynch2013} 
	Lynch R.~S., Bank North Celestial Cap Survey Collaborations, 2013, IAUS, 291, 41

	\bibitem[\protect\citeauthoryear{Lyne \& Graham-Smith}{2006}]{LGS2006} 
	Lyne A. G., Graham-Smith F., 2006, Pulsar astronomy, 3rd ed., Cambridge University Press, Cambridge, UK

	\bibitem[\protect\citeauthoryear{Lyon et al.}{2013}]{Lyon2013} 
	Lyon R.~J., Brooke J.~M., Knowles J.~D., Stappers B.~W., 2013, preprint (arXiv:1307.8012)

	\bibitem[\protect\citeauthoryear{Manchester et al.}{2001}]{Manchester2001} 
	Manchester R.~N., et al., 2001, MNRAS, 328, 17 

	\bibitem[\protect\citeauthoryear{Manchester et al.}{2005}]{Manchester2005} 
	Manchester R.~N., Hobbs G.~B., Teoh A., Hobbs M., 2005, ApJ, 129, 1993 

	\bibitem[\protect\citeauthoryear{McLaughlin \& Cordes}{2003}]{McLaughlin2003}
	McLaughlin M.~A., Cordes J.~M., 2003, ApJ, 596, 982 

	\bibitem[\protect\citeauthoryear{Rojas}{1996}]{Rojas1996} 
	Rojas R., 1996, Neural Networks - A Systematic Introduction, Springer-Verlag

	\bibitem[\protect\citeauthoryear{Smits et al.}{2009}]{Smits2009}
	Smits R., Kramer M., Stappers B., Lorimer D.~R., Cordes J., Faulkner A., 2009, A\&A, 493, 1161 

	\bibitem[\protect\citeauthoryear{Zhu et al.}{2014}]{Zhu2013}
	Zhu W.~W., et al., 2014, ApJ, 781, 117 
\end{thebibliography}
\end{document}